\documentclass[aps,prb,twocolumn,showpacs,floats]{revtex4}
\usepackage{amssymb}

\usepackage{amsmath}
\usepackage{psfig}
\usepackage{bm}
\usepackage{hyperref}

\input{tcilatex}

\begin{document}

\title{Effect of environmental spins on Landau-Zener transitions}
\author{D. A. Garanin$^{\dagger }$, R. Neb$^{\ddagger }$, and R. Schilling$%
^{\ddagger }$}
\affiliation{\mbox{$^{\dagger}$Department of Physics and Astronomy, Lehman College, City
University of New York,} \\ \mbox{250 Bedford Park Boulevard
West, Bronx, New York 10468-1589, U.S.A.} }
\affiliation{$^{\ddagger}$Institut f\"{u}r Physik, Johannes
Gutenberg-Universit\"{a}t, D-55099 Mainz, Germany}
\date{\today}

\begin{abstract}
Landau-Zener (LZ) transitions of a two-level system (e. g., electronic spin
in molecular magnets) coupled to one or many environmental spins (e. g.,
nuclear spins) are studied. For rather general interactions the LZ problem
is reduced to that of a Landau-Zener grid. It is shown analytically that
environmental spins initially in their ground state do not influence the
staying probability $P$. This changes if they are prepared in a statistical
ensemble. For a more specific model with environmental spins in a transverse
field, LZ transitions are studied in the case of well-separated resonances
in the LZ grid. The full evolution of the system is described as a
succession of elementary transitions at avoided crossings and free evolution
between them. If the environmental spins are strongly coupled to the central
spin, their effect on $P$ is weak. In other cases LZ transitions are
strongly suppressed and $P$ is decreasing very slowly with the sweep-rate
parameter $\varepsilon \varpropto 1/v,$ $v$ being the energy sweep rate.
\end{abstract}
\maketitle


\section{Introduction}

Spin tunneling in molecular magnets\cite
{frisartejzio96prl,heretal96epl,thoetal96nat} (see Ref.\ %
\onlinecite{chutej98book} for a review) is strongly affected by the
dipole-dipole interaction (DDI) between the spins $S$ of magnetic molecules
(henceforth electronic spins or simply spins) and by the interaction of the
latter with nuclear spins $I$ (see, e.g., Refs.\ %
\onlinecite{harpolvil96,kuboetal02prb,gotoetal03prb}). Since the tunnel
splitting $\Delta $ is typically rather small, DDI and interaction with
nuclear spins create an energy bias $W$ on electronic spins that can largely
exceed $\Delta $ and thus bring them in or off tunneling resonance. As the
bias due to the DDI depends on the instantaneous configuration of all
electronic spins, spin tunneling and relaxation in molecular magnets is a
complicated many-body phenomenon. Also coupling to nuclear spins renders the
problem a many-body character, especially if the coupling to electronic
spins is strong enough and thus tunneling transitions of the electronic
spins strongly perturb the nuclear subsystem.

There was a big excitement about the experimentally observed slow
nonexponential relaxation at low temperatures in molecular magnets\cite
{weretal99prl,wersesgat99epl} and a great effort to understand it
theoretically using the concept of ``tunneling window''\cite
{cucetal99epjb,alofer01prl,feralo03prl,feralo05prb} defined by $\left|
W\right| \lesssim \Delta $. The role of nuclear spins is believed to be in
creating an effective time-dependent quasirandom bias on electronic spins
that can bring them into the tunneling window and thus allow them to relax
via tunneling.\cite{prosta98prl} However, a similar effect arises via the
DDI due to tunneling of other spins.

Another kind of experiments on molecular magnets is the so-called
Landau-Zener (LZ) experiment\cite{werses99science,weretal00epl} in which the
external magnetic field is time-linearly swept through the resonance and the
magnetization change resulting from spin transitions is monitored. The
asymptotic staying probability in the standard LZ effect is given by\cite
{lan32,zen32,maj32,stu32}
\begin{equation}
P=e^{-\varepsilon },\qquad \varepsilon \equiv \frac{\pi \Delta ^{2}}{2\hbar v%
},  \label{PLZ}
\end{equation}
where $v=\dot{W}=\mathrm{const}$ is the energy sweep rate (contributions of
the four authors are reviewed in Ref.\ \onlinecite{gianik05ufn}). Relating $%
P $ to the observed magnetization change allowed to extract the splitting $%
\Delta $ for the molecular magnet Fe$_{8}.$\cite{werses99science} The
results were in accord with theoretical predictions that the dependence of $%
\Delta $ on a magnetic field along the hard axis shows periodic suppression
of tunneling. \cite{garg93epl} In addition, experimental results for the
effective splitting $\Delta ^{\mathrm{eff}}\left( \varepsilon \right) $ have
demonstrated the existence of an isotope effect originating from different
nuclear spins of the investigated Fe$_{8}$ isotopes.\cite{weretal00epl} A
remarkable property of the LZ effect with linear sweep is independence of
the result for $P$ of the damping of quantum levels\cite{akusch92pra} that
can exceed $\Delta .$ Another remarkable feature is that in the fast-sweep
limit $\varepsilon \ll 1$ the lowest-order result $P\cong 1-\varepsilon $ is
robust with respect to the effects of the interaction discussed below. This
allowed successful interpretation of the experiments at fast sweep in terms
of the standard LZ effect, except for the isotope effect. \cite{weretal00epl}

In the LZ setting, all spins go through the tunneling window at some moment
of time, in spite of the internal bias due to the DDI and nuclear spins.
However, the internal bias has a nontrivial time dependence that makes the
total energy sweep time-nonlinear and drastically changes the LZ effect. In
the quantum-mechanical language, crossing of two energy levels in the
one-particle LZ effect transforms into a series of crossings of many energy
levels in the many-body case (see, e.g., Fig.\ 1 of Ref.\ %
\onlinecite{gar03prb}, Fig.\ 2 of Ref.\ \onlinecite{garsch05prb} or figures
in Refs.\ \onlinecite{wil88jpa,wil89jpa}, where many-body LZ effect is
described by random matrices). The result of multiple quantum transitions at
all crossings is in most cases to increase the staying probability $P$ for
the initial quantum state, compared to the non-interacting case. Especially
strong increase of $P$ (that is, the reduction of the transition probability
$1-P)$ takes place for the ferromagnetic coupling between the electronic
spins. DDI can be considered as a mixture of ferro- and antiferromagnetic
interactions, the former dominating for the crystals elongated in the
direction along the easy axis for electronic spins, that is usually the case.

Analytical and numerical solution of the many-body LZ problem is in general
a formidable task. In particular, a full quantum-mechanical solution is
prohibitive because of a huge Hilbert space, $2^{N}$ for $N$ spins 1/2.
Quantum mean-field approximation (MFA)\cite{hamraemiysai00prb,gar03prb}
makes the problem tractable numerically, since the size of the numerical
problem is only $N.$ A toy model of the many-body LZ effect, $N$ spins all
coupled with the same strength with each other, \cite{gar03prb} also leads
to a numerical problem of size $N,$ that allows to accurately study the
effect of many crossings and to test the validity of the MFA. Probability
for \emph{all} spins to remain in their initial states (that is not related
to the magnetization change) in the LZ model with pairwise interaction
between spins can be calculated analytically. \cite{ostvol06prbrc}

Effect of realistic interactions such as DDI on the LZ effect could only be
studied in the fast-sweep limit, where a perturbative analytical solution at
order $\varepsilon ^{2}$ can be found.\cite{garsch05prb} This allowed to
theoretically explain the experimentally observed onset of the many-body
regime at slower sweeps.\cite{weretal00epl} Up to now no way to
theoretically study the effect of DDI on the LZ effect in molecular magnets
at arbitrary sweep rate could be proposed.

The interaction with nuclear spins strongly reduces the effect of DDI on the
LZ effect that otherwise would be too strong. In Mn$_{12},$ there are 12 Mn
nuclear spins $I=5/2$ in each molecule, coupled to the electronic spin $S$
via the contact hyperfine interaction. Since for nuclear spins this
interaction is much stronger than other interactions such as the Zeeman
interaction with the external field and the nuclear quadrupole interaction,
one can neglect the latter. Then nuclear spins do not have any dynamics and
simply act on the electronic spins as frozen-in random bias fields. The
ensuing inhomogeneous broadening of tunneling resonances strongly reduces
the influence of spin-spin interactions on the LZ effect.\cite{garsch05prb}

In fact, there are a lot of other atoms in magnetic molecules, mainly
hydrogen atoms, that possess nuclear spins. These nuclear spins interact
with the electronic spins via the nuclear dipole-dipole interaction (NDDI).
As the nuclear magnetic moment is by a factor of 10$^{3}$ smaller than the
electronic magnetic moment, NDDI is much weaker than DDI. However, because
of a large number of nuclear spins (120 protons in a Fe$_{8}$ molecule\cite
{weretal00prl}) their cumulative influence on the electronic spins is
appreciable. In contrast to Mn nuclear spins in Mn$_{12},$ interaction of
the remote nuclear spins with the electronic spins is not necessarily
dominating, hence these nuclear spins should have their own dynamics and the
field they produce on the electronic spins is not frozen in.

For a rather general electronic-nuclear spin Hamiltonian, a perturbative
approach up to second order in the bare tunneling matrix element is used in
Ref. \onlinecite{sinpro03prb} which does yield the dependence on the nuclear
spins which are described as fluctuating fields.\cite{prosta98prl}
Restricting to a simplier model, a renormalized splitting obtained by an
instanton technique is used to calculate the transition probability up to
second order. The result \cite{sinpro03prb} coincides with that for the
fluctuating fields, i.e., no isotope effect exists in this approach. A
similar approach by eliminating the nuclear spins within a coherent spin
path integral formalism for the electronic spins has been performed in Ref. %
\onlinecite{garg95prl} to predict suppression of macroscopic quantum
coherence. Since this was done without a sweeping field, no predictions for
the LZ probability can be deduced. A purely numerical solution of the
problem has been presented in Ref.\ \onlinecite
{sindob04prb}.

The aim of the present paper is to clarify the role of nuclear spins
possessing their own dynamics on the basis of standard quantum mechanics
with no usage of stochastic arguments or functional techniques. We will be
using the same simplest model as Ref.\ \onlinecite{sinpro03prb}, an
electronic spin coupled to several nuclear spins, that also experience an
effective field in the perpendicular direction and are initially in the
thermally equilibrium state. Our approach consists of two steps. First we
prove analytically that the LZ problem for a rather general
electronic-nuclear spin Hamiltonian can be exactly mapped to that of a
Landau-Zener grid consisting of two families of parallel ascending and
descending lines vs the energy bias $W.$ This allows one to obtain some
general predictions without specifying the explicit form of the nuclear spin
interactions. Then, in the second step, we construct an analytical solution
based on the standard LZ scattering matrix at every elementary crossing of
the system's energy levels including quantum-mechanical phases.

Problems of this kind have been considered previously.\cite
{demosh68jetp,harpri94pra,har97pra,sinpro03prb} Demkov and Osherov have
shown that in the case of one level crossing several energy levels, the
staying probability after all crossings is a product of all elementary
staying probabilities.\cite{demosh68jetp} Noteworthy is that this conclusion
is valid even in the case of crossings that are \emph{not well separated}
with each other. In the case of \emph{several} parallel lines crossing
another set of parallel lines, Brundobler and Elser made a conjecture that
the same result, multiplication of probabilities, holds for every ascending
line.\cite{bruels93jpa} Refs.\ \onlinecite{usu97prb,shy04pra} present
derivations of the Demkov-Osherov-Brundobler-Elser (DOBE) formula with
different methods. On the other hand, the probability of transition to upper
ascending lines is exactly zero since it would require evolution in a
negative time direction (the so-called no-go theorem\cite
{usu97prb,sin04jpa,volost04jpb}).

For the probabilities of transitions to lower ascending lines and to
descending lines, no general analytical result exists. Since there are
different ``trajectories'' through different intermediate levels on the way
between the initial and final levels, transition probabilities oscillate
because of quantum-mechanical interference (see, e.g., Ref.\ %
\onlinecite{gar04prb}). As in most cases these oscillations should be
averaged out because of the distribution of system's parameters, an
incoherent formalism\cite{harpri94pra,sinpro03prb} neglecting nondiagonal
density matrix coefficients has been used.

What concerns the problem with nuclear spins that we are considering here,
one of our results is that nuclear spins initially in their ground state
does not influence the asymptotic staying probability $P$ for the electronic
spin. On the other hand, here is a substantial increase of $P$ if the
nuclear spins are initially in their excited states. In the latter case and
more generally for nuclear spins at finite temperatures, the staying
probability exhibits an isotope effect. However, the effect disappears again
if nuclear spins are strongly coupled to the electronic spin, as well as for
arbitrary coupling constant in the fast-sweep limit.

The paper is organized as follows. Sec.\ \ref{Sec-General} introduces a
general model for a central spin 1/2 coupled to environmental spins and
undergoing a LZ transition. The Hamiltonian of environmental spins can be
arbitrary, including their interaction with each other. It is shown how the
Hamiltonian of the system can be diagonalized with respect to the
environmental spins, that leads to the Landau-Zener grid. In Sec.\ \ref
{Sec-Basic} a particular model is introduced, the model with one nuclear
spin $I$ in an arbitrarily directed effective field. This model can be
easily diagonalized by choosing an appropriate frame for the nuclear spin
having the $z^{\prime }$ axis along the total field. In Sec.\ \ref
{Sec-Transitions} the staying probability $P$ on the LZ grid is calculated
for well-separated resonances by multiplying LZ scattering matrices of
elementary avoided crossings. In Sec.\ \ref{Sec-Many-incoh} the model is
generalized for $N$ equivalent nuclear spins $I$. This model can be reduced
to the model with one nuclear spin as the states of the nuclear subsystem
can be classified in the total nuclear spin $I_{\mathrm{tot}}$ that is
dynamically conserved. In this section the effect of quantum-mechanical
phase oscillations is considered. Averaging over initial thermal
distribution of nuclear spins and over different $I_{\mathrm{tot}}$ leads to
only partial averaging out of the oscillations for the simulated system
sizes. Here the more realistic incoherent approximation is introduced that
leads to smooth dependences $P(\varepsilon ).$ The concluding section
contains a discussion of the two main cases: Mn$^{55}$ nuclear spins in the
Mn$_{12}$ molecular magnet and nuclear spins of the protons.

\section{General formulation}

\label{Sec-General}

Tunneling of an electronic spin $S$ of a magnetic molecule under the barrier
can be considered in the approximation of two resonant levels as dynamics of
a pseudospin 1/2 coupled to $N$ environmental spins, say, nuclear spins $I$.
The Hamiltonian has the form
\begin{equation}
\hat{H}(t)=-\frac{1}{2}W(t)\sigma _{z}-\frac{1}{2}\Delta \sigma _{x}-\sigma
_{z}\hat{V}\left( \left\{ \mathbf{I}\right\} \right) +\hat{H}_{\mathrm{nuc}%
}\left( \left\{ \mathbf{I}\right\} \right) ,  \label{HamGen}
\end{equation}
where $\left\{ \mathbf{I}\right\} \equiv \mathbf{I}_{1},\ldots ,\mathbf{I}%
_{N}$, and $\sigma _{z}$, $\sigma _{x}$ are Pauli matrices. We assume a
time-linear energy sweep $W(t)=vt$ with $v=\mathrm{const}.$ The third term
here describes the contact hyperfine coupling or dipole-dipole interaction
between the electronic spin and nuclear spins. The fourth term is the
nuclear Hamiltonian that can also contain interaction between nuclear spins.

Let us introduce basis states
\begin{equation}
\left| \xi ;\left\{ m_{I}\right\} \right\rangle =\left| \xi \right\rangle
\otimes \left| \left\{ m_{I}\right\} \right\rangle ,\qquad \xi =\pm 1,\quad
m_{I}^{(i)}=-I,\ldots ,I,
\end{equation}
$\left\{ m_{I}\right\} \equiv m_{I}^{(1)},\ldots ,m_{I}^{(N)},$ that form an
eigenbasis of $\sigma _{z}$ and $I_{z}^{(i)}$:
\begin{eqnarray}
\sigma _{z}\left| \xi ;\left\{ m_{I}\right\} \right\rangle &=&\xi \left| \xi
;\left\{ m_{I}\right\} \right\rangle ,  \notag \\
I_{z}^{(i)}\left| \xi ;\left\{ m_{I}\right\} \right\rangle
&=&m_{I}^{(i)}\left| \xi ;\left\{ m_{I}\right\} \right\rangle .
\end{eqnarray}
The Hamiltonian matrix
\begin{equation}
\mathbb{H}(t)\equiv \left\{ \left\langle \xi ^{\prime };\left\{
m_{I}^{\prime }\right\} \left| \hat{H}(t)\right| \xi ;\left\{ m_{I}\right\}
\right\rangle \right\}
\end{equation}
can be represented in the following block form corresponding to the two
electronic states:
\begin{equation}
\mathbb{H}(t)=\left(
\begin{array}{cc}
\mathbb{H}_{--}(t) & \mathbb{H}_{-+} \\
\mathbb{H}_{+-} & \mathbb{H}_{++}(t)
\end{array}
\right) ,  \label{HamGenMatr}
\end{equation}
where $\mathbb{H}_{+-}=\mathbb{H}_{-+}=-\left( \Delta /2\right) \mathbb{I}$,
$\mathbb{I}$ being a unit matrix in nuclear indices, and
\begin{equation*}
\mathbb{H}_{\pm \pm }(t)=\mp \frac{1}{2}W(t)\mathbb{I}\mp \mathbb{V}+\mathbb{%
H}_{\mathrm{nuc}}.
\end{equation*}
Here the $\left( 2I+1\right) ^{N}\times \left( 2I+1\right) ^{N}$ matrices
with respect to nuclear indices are defined by
\begin{equation}
\mathbb{V}\equiv \left\{ \left\langle \left\{ m_{I}^{\prime }\right\} \left|
\hat{V}\right| \left\{ m_{I}\right\} \right\rangle \right\} ,
\end{equation}
etc.

Since $\mathbb{H}_{\pm \pm }$ is Hermitean, it can be diagonalized by a
unitary transformation matrix $\mathbb{U}_{\pm }$%
\begin{equation}
\mathbb{U}_{\pm }\mathbb{H}_{\pm \pm }(t)\mathbb{U}_{\pm }^{-1}=\mathbb{D}%
_{\pm }(t),  \label{DMatr}
\end{equation}
$\mathbb{D}_{\pm }(t)$ being diagonal. Applying the block unitary
transformation
\begin{equation}
\mathbb{U=}\left(
\begin{array}{cc}
\mathbb{U}_{-} & \mathbb{O} \\
\mathbb{O} & \mathbb{U}_{+}
\end{array}
\right)
\end{equation}
to Eq.\ (\ref{HamGenMatr}), $\mathbb{O}$ being zero matrix, one obtains the
transformed Hamiltonian matrix
\begin{equation}
\mathbb{H}^{\prime }(t)=\left(
\begin{array}{cc}
\mathbb{D}_{-}(t) & \mathbb{V}^{\prime } \\
\mathbb{V}^{\prime \dagger } & \mathbb{D}_{+}(t)
\end{array}
\right) ,  \label{HamGenMatrpr}
\end{equation}
where
\begin{equation}
\mathbb{V}^{\prime }=-\left( \Delta /2\right) \mathbb{U}_{-}\mathbb{U}%
_{+}^{-1},\qquad \mathbb{V}^{\prime \dagger }=-\left( \Delta /2\right)
\mathbb{U}_{+}\mathbb{U}_{-}^{-1}.  \label{VMatr}
\end{equation}

In the new basis $\left| \xi ;k\right\rangle $, there are two sets of
nuclear states for the electronic spin up and down, $\xi =\pm 1$. Diagonal
elements of the diagonal matrices $\mathbb{D}_{\pm }(t)$ have the form
\begin{equation}
D_{\xi ,k}(t)=-\frac{\xi }{2}W(t)+d_{\xi ,k},
\end{equation}
$k=1,\ldots ,\left( 2I+1\right) ^{N},$ sorted so that $d_{\xi ,k}$ increases
with $k.$ Elements of $\mathbb{V}^{\prime }$ are
\begin{equation}
V_{kk^{\prime }}^{\prime }=-\left( \Delta /2\right) \left( \mathbb{U}_{-}%
\mathbb{U}_{+}^{-1}\right) _{kk^{\prime }}.  \label{Vkkpr}
\end{equation}
Avoided crossing of $\left( 2I+1\right) ^{N}$ ascending $\left( \xi
=-1\right) $ lines with $\left( 2I+1\right) ^{N}$ descending $\left( \xi
=1\right) $ lines forms the Landau-Zener grid shown in Fig.\ \ref
{Fig-LZ-grid} in a particular case of one nuclear spin $I=1$. In general,
all crossings are avoided crossings with splittings $\Delta _{kk^{\prime
}}=\left| 2V_{kk^{\prime }}^{\prime }\right| ,$ as defined by Eq.\ (\ref
{Vkkpr}). Crossing of $\left| 1;k\right\rangle $ with $\left| -1;k^{\prime
}\right\rangle $ occurs at $D_{+,k}(t)=D_{-,k^{\prime }}(t)$ or
\begin{equation}
W(t)=W_{kk^{\prime }}\equiv d_{+,k}-d_{-,k^{\prime }}.
\end{equation}

If the system is prepared at $t=-\infty $ in the state $\left|
-1,k\right\rangle ,$ then the probability $P_{kk}$ to remain in this state
is given by the DOBE formula as the product of staying probabilities at all
crossings:\cite{demosh68jetp,bruels93jpa,usu97prb,shy04pra}
\begin{eqnarray}
P_{kk} &=&\exp \left[ -\varepsilon \sum_{k^{\prime }}\left( \Delta
_{kk^{\prime }}/\Delta \right) ^{2}\right]  \notag \\
&=&\exp \left[ -\varepsilon \sum_{k^{\prime }}\left| \left( \mathbb{U}_{-}%
\mathbb{U}_{+}^{-1}\right) _{kk^{\prime }}\right| ^{2}\right] .  \label{Pkk0}
\end{eqnarray}
With the use of
\begin{eqnarray}
&&\sum_{k^{\prime }}\left| \left( \mathbb{U}_{-}\mathbb{U}_{+}^{-1}\right)
_{kk^{\prime }}\right| ^{2}  \notag \\
&=&\sum_{k^{\prime }}\left( \mathbb{U}_{-}\mathbb{U}_{+}^{-1}\right)
_{kk^{\prime }}\left( \mathbb{U}_{-}\mathbb{U}_{+}^{-1}\right) _{k^{\prime
}k}^{\dagger }  \notag \\
&=&\left( \mathbb{U}_{-}\mathbb{U}_{+}^{-1}\mathbb{U}_{+}\mathbb{U}%
_{-}^{-1}\right) _{kk}=\left\{ \mathbb{I}\right\} _{kk}=1
\end{eqnarray}
one obtains
\begin{equation}
P_{kk}=e^{-\varepsilon },  \label{Pkk}
\end{equation}
the same as Eq.\ (\ref{PLZ}). This result shows that $P_{kk}$ does not
depend on nuclear spins, i. e., there is no isotope effect. However, since
we are interested in transitions of the electronic spin alone, the relevant
quantity is
\begin{equation}
P=\sum_{kk^{\prime }}n_{k}P_{kk^{\prime }},
\end{equation}
where $n_{k}$ are populations of the nuclear levels in the initial state.
This formula assumes that initially nuclear spins are in a state described
by a diagonal density matrix, e.g., a thermal equilibrium state. With
account of the no-go theorem\cite{usu97prb,sin04jpa,volost04jpb} and Eq.\ (%
\ref{Pkk}), the result reduces to
\begin{equation}
P=e^{-\varepsilon }+\sum_{k>k^{\prime }}n_{k}P_{kk^{\prime }}.
\label{PnucGeneral}
\end{equation}
If all nuclear spins are initially in the ground state ($n_{k}=\delta
_{k,1}),$ there are no terms in the sum and the standard LZ result is
reproduced. Otherwise $P$ exhibits an isotope effect and increases above $%
e^{-\varepsilon },$ i.e., coupling to nuclear spins is hampering spin
transitions. For not too low temperatures, nuclear spins are equidistributed
in the initial state, i. e., $n_{k}=\left( 2I+1\right) ^{-N}.$

The effect of hampering spin transitions is especially strong at slow sweep.
Indeed, at slow sweep in the case of a standard LZ effect, the system
practically follows the lower adiabatic level, so that the probability $P$
of a transition to the upper adiabatic level (i. e., of staying on the
accending diabatic level) is exponentially small. For the model with nuclear
spins, the starting level of the system is in general not the lowest level,
because nuclear spins are thermally distributed over their energy levels. In
addition, splittings $\Delta _{kk^{\prime }}$ are widely distributed, so
that there are very small splittings and the adiabatic limit is practically
never reached. As a result, there are a lot of transitions between ascending
and descending levels in both directions, so that the electronic spin
performs a complicated motion on the Landau-Zener grid.

Consideration in this section suggests that interaction between $N$ nuclear
spins does not change the situation qualitatively. Indeed, Eq.\ (\ref
{HamGenMatrpr}) has the same form with and without interaction, so that the
topology of the LZ grid is not affected. Numerical results of Ref.\
\onlinecite
{sindob04prb} show only a moderate effect of interaction between nuclear
spins. Anyway, all feasible types of interactions between nuclear spins,
such as the direct dipole-dipole interaction and indirect interactions via
the electronic spins, are much weaker than the interaction of nuclear spins
with the electronic spin and interaction of nuclear spins with the external
magnetic field or with the gradients of the microscopic electric field\cite
{kuboetal02prb,gotoetal03prb} due to the quadrupole moment of nuclei.

In the next section we will consider our basic model of an electronic spin
interacting with one nuclear spin in an effective field that can be easily
diagonalized. Generalization for the case of many nuclear spins will be done
later in Sec.\ \ref{Sec-Many-incoh}.

\section{The model Hamiltonian and its diagonalization}

\label{Sec-Basic}

Consider the Hamiltonian for an electronic spin coupled to a single
environmental spin
\begin{equation}
\hat{H}=-\frac{1}{2}W(t)\sigma _{z}-\frac{1}{2}\Delta \sigma _{x}-A\sigma
_{z}I_{z}-\Lambda _{z}I_{z}-\Lambda _{x}I_{x}  \label{HamSI}
\end{equation}
that is a particular form of Eq.\ (\ref{HamGen}). Terms with $A^{\prime
}\sigma _{z}I_{x},$ $A^{\prime \prime }\sigma _{z}I_{y},$ and $\Lambda
_{y}I_{y}$ can be added to the Hamiltonian but such terms can be eliminated
by choosing its own system of axes $x^{\prime },y^{\prime },z^{\prime }$ for
the nuclear spin. $A$ stands for the contact hyperfine interaction or for
NDDI between the electronic spin and protons, as said above. In Eq.\ (\ref
{HamSI}) $\Lambda _{z}$ and $\Lambda _{x}$ are energy-dimensional components
of the field acting on nuclear spins, e.g., an external magnetic field, $%
\Lambda _{\alpha }=g_{n}\mu _{n}H_{\alpha },$ where $\alpha =x,y,$ $\mu _{n}$
is the nuclear magneton, and $g_{n}$ is the nuclear Land\'{e} factor. A Mn$%
^{55}$ nucleus in the molecular magnet Mn$_{12}$ has $I=5/2$ and the
magnetic moment $\mu _{\mathrm{Mn}^{55}}=3.45\mu _{n},$ so that $g_{n,%
\mathrm{Mn}^{55}}\simeq 3.45/I\simeq 1.38.$ \ A proton has $I=1/2$ and the
magnetic moment $\mu _{p}=2.79\mu _{n},$ so that $g_{n,p}\simeq 2.79/I\simeq
5.58.$ For the contact hyperfine interaction with the Mn$^{55}$ nuclear
spins in Mn$_{12}$ one has $A/k_{\mathrm{B}}=0.02$ K. This interaction is
very strong and equivalent to a magnetic field of 40 T applied to nuclear
spins.

For the description of the states of the nuclear spin it is convenient to
use the basis of its eigenstates corresponding to the states $\left|
\downarrow \right\rangle $ and $\left| \uparrow \right\rangle ,$ or $\xi
=\pm ,$ of the electronic spin. Fixing the electronic spin in the state $\xi
$ creates an effective field $\xi A_{z}\mathbf{e}_{z}$ on the nuclear spin
then gives the effective nuclear Hamiltonian
\begin{equation}
\hat{H}_{\xi ,n,\mathrm{eff}}=-\mathbf{F}_{\xi }\cdot \mathbf{I,}
\label{Hneff}
\end{equation}
where
\begin{equation}
\mathbf{F}_{\xi }=\left( \xi A+\Lambda _{z}\right) \mathbf{e}_{z}+\Lambda
_{x}\mathbf{e}_{x}  \label{HfilednDef}
\end{equation}
is the total field acting on the nuclear spin. This Hamiltonian can be
diagonalized by choosing the $z_{\xi }^{\prime }$ axis for the nuclear spin
in the direction of $F_{\xi },$ i.e.,
\begin{equation}
\mathbf{e}_{z_{\xi }^{\prime }}=\frac{\xi A+\Lambda _{z}}{F_{\xi }}\mathbf{e}%
_{z}+\frac{\Lambda _{x}}{F_{\xi }}\mathbf{e}_{x},  \label{ezprDef}
\end{equation}
where
\begin{equation}
F_{\xi }=\sqrt{\left( \xi A+\Lambda _{z}\right) ^{2}+\Lambda _{x}^{2}}.
\label{OmeganDef}
\end{equation}
The vectors $\mathbf{e}_{z_{\xi }^{\prime }}$ are rotated away from $\mathbf{%
e}_{z}$ by the angles
\begin{equation}
\beta _{\xi }=\arccos \frac{\xi A+\Lambda _{z}}{F_{\xi }}.  \label{betaxiDef}
\end{equation}
For $\Lambda _{x}\ll A,\Lambda _{z}$ and $\Lambda _{z}<A,$ $\beta _{+}$ is
close to 0 and $\beta _{-}$ is close to $\pi .$ For $\Lambda _{x}\ll
A,\Lambda _{z}$ and $A<\Lambda _{z},$ both $\beta _{+}$ and $\beta _{-}$ are
close to 0. For the transverse nuclear axes we define
\begin{equation}
\mathbf{e}_{y_{\xi }^{\prime }}=\mathbf{e}_{y},\qquad \mathbf{e}_{x_{\xi
}^{\prime }}=\left[ \mathbf{e}_{y}\times \mathbf{e}_{z_{\xi }^{\prime }}%
\right] =\frac{\xi A+\Lambda _{z}}{F_{\xi }}\mathbf{e}_{x}-\frac{\Lambda _{x}%
}{F_{\xi }}\mathbf{e}_{z}.
\end{equation}

The corresponding basis set of states has the form
\begin{equation}
\left| \xi ,m_{I}\right\rangle =\left| \xi \right\rangle \left|
m_{I}\right\rangle _{\xi },  \label{BasisStates}
\end{equation}
where the rotated states $\left| m_{I}\right\rangle _{\xi }$ depend on $\xi .
$ Elements of the Hamiltonian matrix $\mathbb{H}^{\prime }$ in the new
basis, Eq.\ (\ref{HamGenMatrpr}), are defined by
\begin{equation}
H_{\xi ,m_{I};\xi ^{\prime },m_{I}^{\prime }}^{\prime }=\left. _{\xi
}\left\langle m_{I}\right| \left\langle \xi \right| \hat{H}\left| \xi
^{\prime }\right\rangle \left| m_{I}^{\prime }\right\rangle _{\xi ^{\prime
}}\right. .  \label{VDef}
\end{equation}
With the help of $\mathbf{I=}I_{x_{\xi }^{\prime }}\mathbf{e}_{x_{\xi
}^{\prime }}+I_{y}\mathbf{e}_{y}+I_{z_{\xi }^{\prime }}\mathbf{e}_{z_{\xi
}^{\prime }}$ one obtains
\begin{equation}
H_{\xi ,m_{I};\xi ^{\prime },m_{I}^{\prime }}^{\prime }=E_{\xi ,m_{I}}\delta
_{\xi ^{\prime }\xi }\delta _{m_{I}^{\prime }m_{I}}-\frac{1}{2}\Delta _{\xi
,m_{I},m_{I}^{\prime }}\delta _{\xi ^{\prime },-\xi },  \label{MEHam}
\end{equation}
where
\begin{equation}
E_{\xi ,m_{I}}=-\frac{1}{2}\xi W(t)-F_{\xi }m_{I}  \label{EDef}
\end{equation}
are the diagonal elements of $\mathbb{D}_{\pm }$ of Eq. (\ref{DMatr}).
\begin{equation}
\Delta _{\xi ,m_{I},m_{I}^{\prime }}=\Delta \left. _{\xi }\left\langle
m_{I}\right. \left| m_{I}^{\prime }\right\rangle _{-\xi }\right. ,
\label{Deltammpr}
\end{equation}
are the elements of $\mathbb{V}^{\prime }$ in Eq. (\ref{VMatr}) for $\xi =-1,
$ and the elements of $\mathbb{V}^{\prime \dagger }$ for $\xi =1$. Elements
of the Hamiltonian matrix are placed in order of $\xi $ changing from $-1$
to 1 and $m_{I}$ changing from $I$ to $-I.$

The projector $_{\xi }\left\langle m_{I}\right. \left| m_{I}^{\prime
}\right\rangle _{-\xi }$ can be expressed through the spin rotation matrix
\cite{hecht00book}
\begin{eqnarray}
&&d_{m^{\prime }m}^{\left( I\right) }(\beta )=\left\langle m^{\prime }\left|
e^{-i\beta I_{y}}\right| m\right\rangle   \notag \\
&=&\left[ \frac{(I-m)!(I-m^{\prime })!}{(I+m)!(I+m^{\prime })!}\right]
^{1/2}\left( \cos \frac{\beta }{2}\right) ^{m+m^{\prime }}\left( \sin \frac{%
\beta }{2}\right) ^{m-m^{\prime }}  \notag \\
&&\times \sum_{n}\frac{(-1)^{n}(I+m+n)!}{(I-m-n)!(m+n-m^{\prime })!n!}\left(
\sin \frac{\beta }{2}\right) ^{2n},  \label{dHecht}
\end{eqnarray}
summation going over $\max (0,m^{\prime }-m)\leq n\leq I-m.$ The final
expression is good for $\beta >0.$ In the case $\beta <0$ one should use the
relation $d_{m^{\prime }m}^{\left( J\right) }(\beta )=\left( -1\right)
^{m^{\prime }-m}d_{m^{\prime }m}^{\left( J\right) }(-\beta ).$ For large $I$
numerical implementation of Eq.\ (\ref{dHecht}) leads to precision problems.
In this case it is much more convenient to obtain $d_{m^{\prime }m}^{\left(
I\right) }(\beta )$ numerically by finding eigenstates of the operator $%
I_{z}\cos \beta +I_{x}\sin \beta $ and projecting them on $\left|
m\right\rangle ,$ eigenstates of $I_{z}.$

Using $\left| m_{I}\right\rangle _{\xi }=e^{-i\beta _{\xi }I_{y}}\left|
m_{I}\right\rangle ,$ where $\left| m_{I}\right\rangle $ are the states of
the initial basis, quantized along the $z$-axis, one obtains
\begin{eqnarray}
_{\xi }\left\langle m_{I}\right. \left| m_{I}^{\prime }\right\rangle _{-\xi
} &=&\left\langle m_{I}\left| e^{i\beta _{\xi }I_{y}}e^{-i\beta _{-\xi
}I_{y}}\right| m_{I}^{\prime }\right\rangle =d_{m_{I}m_{I}^{\prime
}}^{\left( I\right) }(\beta )  \notag \\
\beta  &=&-\beta _{\xi }+\beta _{-\xi }  \label{Projd}
\end{eqnarray}
For $\Lambda _{z}=0$ from Eq.\ (\ref{betaxiDef}) one obtains
\begin{equation}
\beta =2\xi \arcsin \frac{A}{F},  \label{betadiff}
\end{equation}
where $F=\sqrt{A^{2}+\Lambda _{x}^{2}},$ and, in Eq.\ (\ref{Deltammpr})
\begin{equation}
\Delta _{\xi ,m_{I},m_{I}^{\prime }}=\Delta \xi ^{m_{I}-m_{I}^{\prime
}}d_{m_{I}m_{I}^{\prime }}^{\left( I\right) }\left( 2\arcsin \frac{A}{F}%
\right) .  \label{Deltammprdmmpr}
\end{equation}

In Eq.\ (\ref{MEHam}) $E_{\xi ,m_{I}}$ describes $2I+1$ ascending ($\xi =-1$%
) and $2I+1$ descending ($\xi =1$) lines. All crossings are avoided
crossings, because all $d_{m_{I}m_{I}^{\prime }}^{\left( I\right) }\neq 0.$
The sum rule
\begin{equation}
\sum_{m^{\prime }=-I}^{I}\left[ d_{m^{\prime }m}^{\left( I\right) }(\beta )%
\right] ^{2}=1
\end{equation}
leads to Eq.\ (\ref{Pkk}) in our particular case.

\begin{figure}[t]
\unitlength1cm
\begin{picture}(11,5.5)
\centerline{\psfig{file=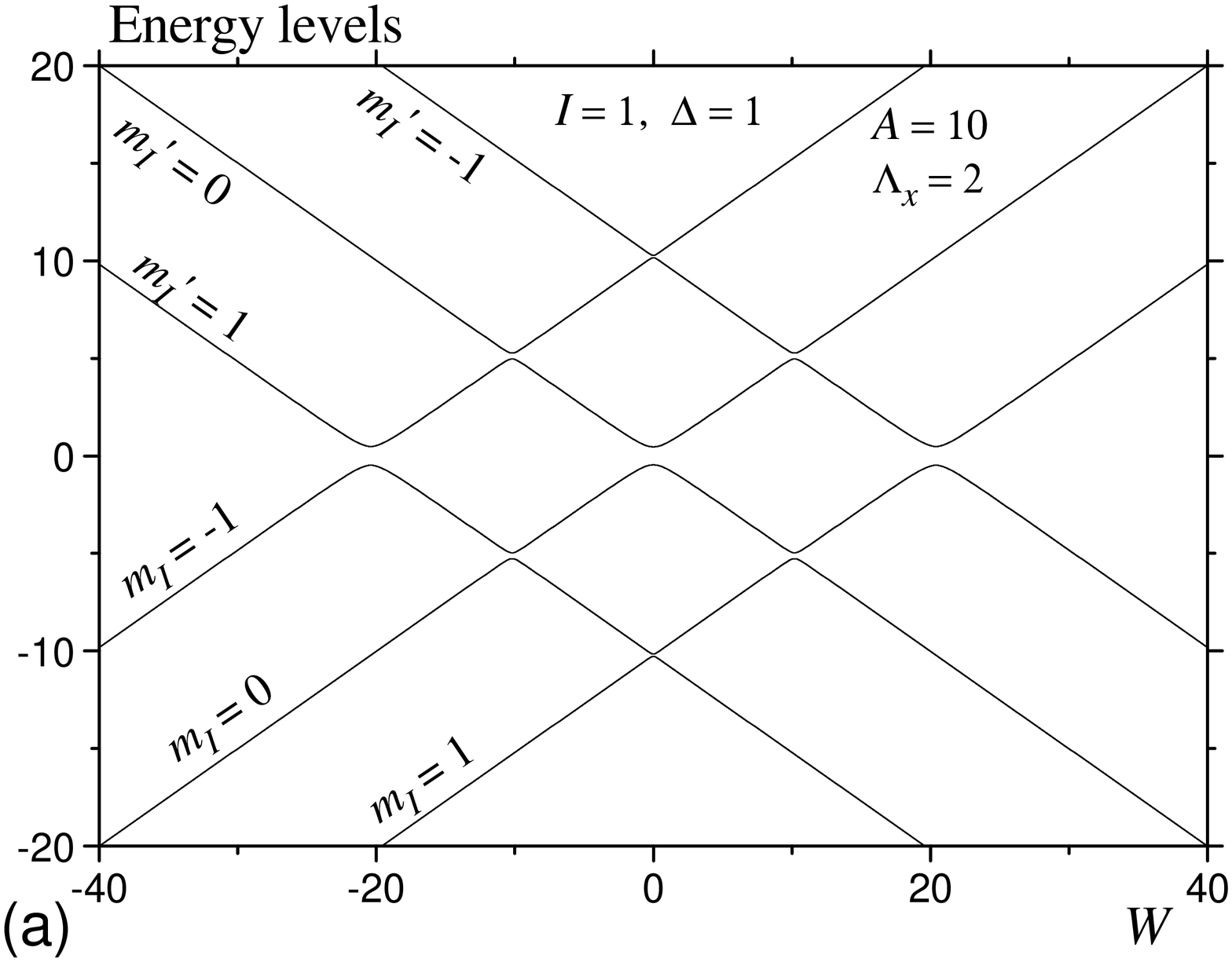,angle=-90,width=8cm}}
\end{picture}
\begin{picture}(11,5.5)
\centerline{\psfig{file=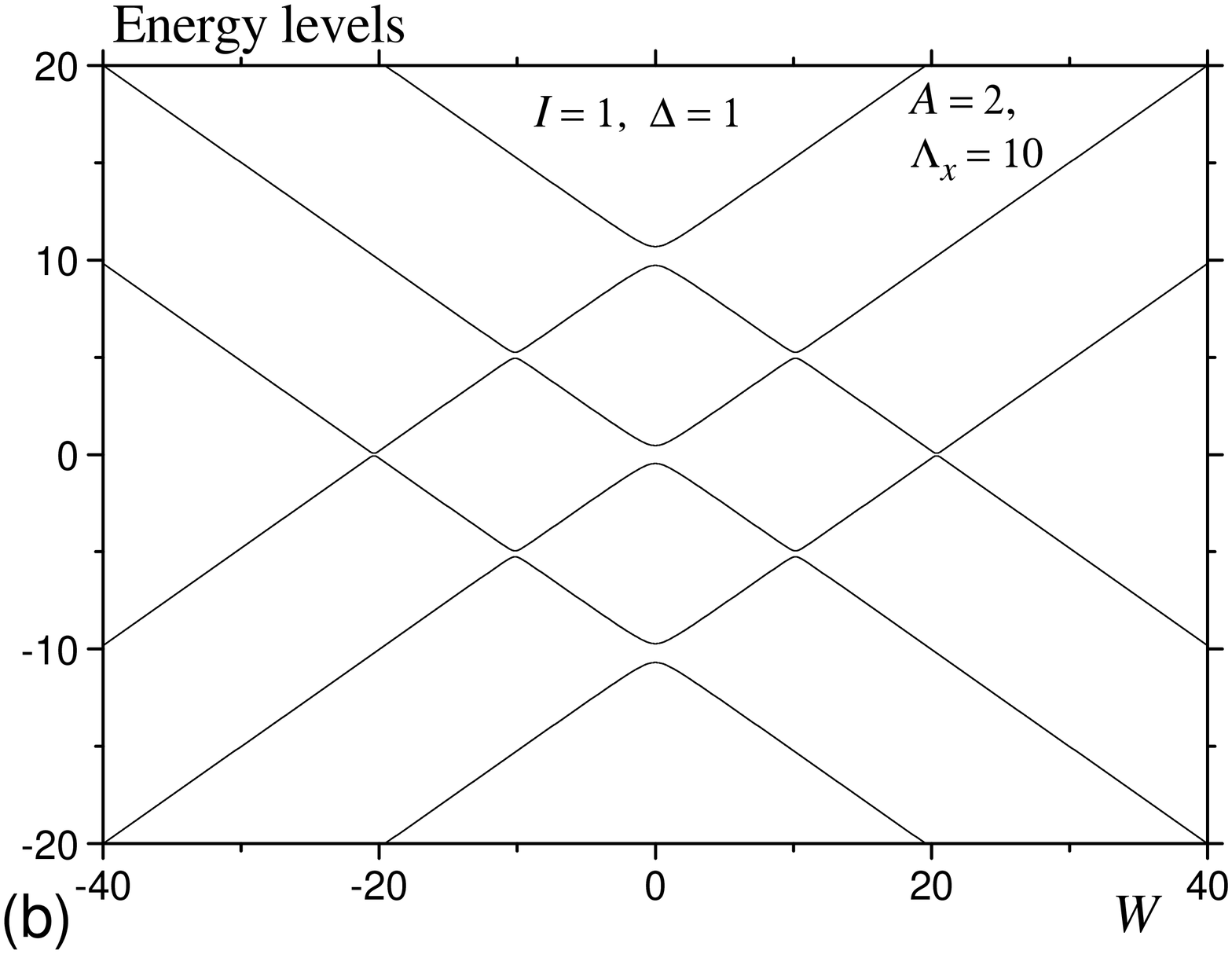,angle=-90,width=8cm}}
\end{picture}
\caption{Landau-Zener grid for one nuclear spin $I=1$ and $\Lambda _{z}=0$:
(a) $A\gg \Lambda _{x}$; (b) $A\ll \Lambda _{x}$;}
\label{Fig-LZ-grid}
\end{figure}
\begin{figure}[t]
\unitlength1cm
\begin{picture}(11,5.5)
\centerline{\psfig{file=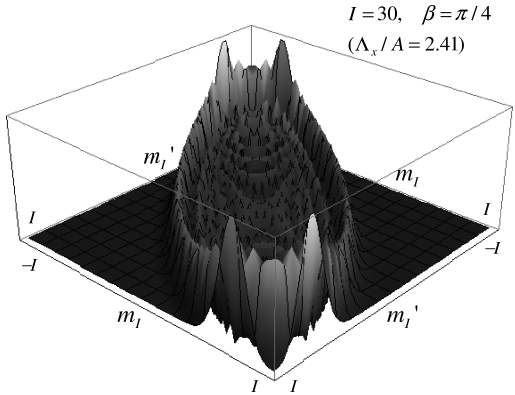,angle=-90,width=8cm}}
\end{picture}
\caption{Squares of splittings, $\left( \Delta _{m_{I}m_{I}}^{\prime
}\right) ^{2}\varpropto \left[ d_{m_{I}m_{I}}^{(I)}(\protect\beta )\right]
^{2}$ on the LZ grid for a large ``nuclear spin'' $I=30$ and $\protect\beta =%
\protect\pi /4$ ($\Lambda _{x}/A=2.41$).}
\label{Fig-Levels-Large_spin}
\end{figure}

For $\Lambda _{z}=0,$ one obtains the matrix $\mathbb{V}^{\prime }$\ \ in
Eq.\ (\ref{HamGenMatrpr}) ($\xi =-1,$ $\xi ^{\prime }=1)$ in the form
\begin{equation}
\mathbb{V}^{\prime }=-\frac{\Delta }{2}\left(
\begin{array}{cc}
\frac{\Lambda _{x}}{F} & \frac{A}{F} \\
-\frac{A}{F} & \frac{\Lambda _{x}}{F}
\end{array}
\right)   \label{VMatrOneHalf}
\end{equation}
for $I=1/2,$
\begin{equation}
\mathbb{V}^{\prime }=-\frac{\Delta }{2}\left(
\begin{array}{ccc}
\frac{\Lambda _{x}^{2}}{F^{2}} & \sqrt{2}\frac{\Lambda _{x}A}{F^{2}} & \frac{%
A^{2}}{F^{2}} \\
-\sqrt{2}\frac{\Lambda _{x}A}{F^{2}} & \frac{\Lambda _{x}^{2}-A^{2}}{F^{2}}
& \sqrt{2}\frac{\Lambda _{x}A}{F^{2}} \\
\frac{A^{2}}{F^{2}} & -\sqrt{2}\frac{\Lambda _{x}A}{F^{2}} & \frac{\Lambda
_{x}^{2}}{F^{2}}
\end{array}
\right)   \label{VMatrOne}
\end{equation}
for $I=1,$ etc. The matrix elements depend only on the relation between $%
\Lambda _{x}$ and $A$ in Eq.\ (\ref{HamSI}). For $\Lambda _{x}\gg A$ the
matrix $\mathbb{V}^{\prime }$ becomes nearly diagonal, that results from $%
\beta _{\xi }\approx \beta _{-\xi }$ thus $\beta \approx 0.$ In the opposite
case $\Lambda _{x}\ll A$ matrix $\mathbb{V}^{\prime }$ becomes nearly
antidiagonal, that results from $\beta =\beta _{-\xi }-\beta _{\xi }\approx
\pi .$ For $\Lambda _{x}\gg A$ level splittings $\Delta _{\xi
,m_{I},m_{I}^{\prime }}=2\left| V_{\xi ,m_{I},m_{I}^{\prime }}^{\prime
}\right| $ are large between the levels having the same $m_{I},$ other
splittings are small, as shown in Fig.\ \ref{Fig-LZ-grid}, (b). This means
that transitions of the electronic spin mostly leave nuclear spins at their
initial levels, a natural result for a small coupling to spins. For $\Lambda
_{x}\ll A$, the maximal splittings are those between the levels with nuclear
indices related by $m_{I}^{\prime }=-m_{I},$ while other splittings are
small, as shown in Fig.\ \ref{Fig-LZ-grid}, (a). In the original basis with
the quantization axis $z,$ this means that the states of nuclear spins do
not change, too, similarly to the other limiting case. Indeed, for $\Lambda
_{x}=0$ there are no transitions of nuclear spins since the latter have no
dynamics. For $\Lambda _{z}\neq 0,$ matrix $\mathbb{V}^{\prime }$ is
cumbersome but has the same properties: Nearly diagonal for $\beta \approx 0$
(small $A)$ and nearly antidiagonal for $\beta \approx \pi $ (large $A).$

For large $I$ the distribution of splittings in the plane $\left(
m_{I},m_{I}^{\prime }\right) ,$ defined by the spin rotation matrix $%
d_{m_{I}m_{I}^{\prime }}^{\left( I\right) }\left( \beta \right) ,$ becomes
more interesting, see Fig.\ \ref{Fig-Levels-Large_spin}. The splittings
become very small outside an ellipse and maximal on the ellipse. Inside the
ellipse, $d_{m_{I}m_{I}^{\prime }}^{\left( I\right) }\left( \beta \right)
\sim 1$ but changes abruptly if the indices change by one. For $\beta =\pi
/2 $ the ellipse becomes a circle of radius $I.$ For $\beta \rightarrow 0$
or $\beta \rightarrow \pi $ the ellipse degenerates into a vertical ($%
m_{I}^{\prime }=m_{I}$) or horizontal ($m_{I}^{\prime }=-m_{I}$) lines,
respectively. For large $I,$ even if $\beta $ is close to 0 ($\Lambda
_{x}/A\gg 1$) or $\pi $ ($\Lambda _{x}/A\ll 1$), there are many crossings
inside the ellipse where splittings are comparable to $\Delta $.

In particular, for $\beta \ll 1$ the matrix $d_{m^{\prime }m}^{\left(
I\right) }(\beta )$ of Eq.\ (\ref{dHecht}) becomes nearly diagonal and can
be expanded directly as
\begin{eqnarray}
d_{m^{\prime }m}^{\left( I\right) }(\beta ) &\cong &\delta _{m^{\prime
}m}-\beta \left\langle m^{\prime }\left| iI_{y}\right| m\right\rangle
\notag \\
&=&\delta _{m^{\prime }m}-\frac{\beta }{2}\left( \delta _{m^{\prime
},m+1}l_{m,m+1}\right.   \notag \\
&&-\left. \delta _{m^{\prime },m-1}l_{m,m-1}\right) ,  \label{dSmallbeta}
\end{eqnarray}
where $l_{m,m+1}=\sqrt{I(I+1)-m(m+1)}.$ This is the case realized for $%
\Lambda _{x}\gg A,$ where from Eq.\ (\ref{betadiff}) with $\xi =-1$ follows $%
\beta \cong -2A/F\cong -2A/\Lambda _{x}$ and in Eq.\ (\ref{Deltammpr})
\begin{eqnarray}
\frac{\Delta _{-1,m_{I},m_{I}^{\prime }}}{\Delta } &=&\delta
_{m_{I}m_{I}^{\prime }}+\frac{A}{\Lambda _{x}}\left( \delta _{m_{I}^{\prime
},m_{I}-1}l_{m_{I},m_{I}-1}\right.   \notag \\
&&-\left. \delta _{m_{I}^{\prime },m_{I}+1}l_{m_{I},m_{I}+1}\right) ,
\label{DeltammLamSmall}
\end{eqnarray}
in accordance with Eqs.\ (\ref{VMatrOneHalf}) and (\ref{VMatrOne}) in this
limit. Similarly, the case $\Lambda _{x}\ll A$ corresponds to $\beta $ close
to $\pi ,$ so that $d_{m^{\prime }m}^{\left( I\right) }(\beta )$ becomes
nearly anti-diagonal. The elements adjacent to the antidiagonal are
proportional to $\left( \Lambda _{x}/A\right) l_{m_{I},m_{I}+1}.$ In both
cases, diagonal or anti-diagonal form of $d_{m^{\prime }m}^{\left( I\right)
}(\beta )$ requires $IA/\Lambda _{x}\ll 1$ or $I\Lambda _{x}/A\ll 1,$
respectively.

\section{Transitions on the Landau-Zener grid}

\label{Sec-Transitions}

As in the general case considered in Sec.\ \ref{Sec-General}, the energy
levels described by Eq.\ (\ref{EDef}) separate into two families, ascending
lines for $\xi =-1$ and descending lines for $\xi =+1.$ There are crossings
between these families at
\begin{equation}
E_{+,m_{I}^{(+)}}(t)=E_{-,m_{I}^{(-)}}(t).
\end{equation}
Each ascending line crosses with $2I+1$ descending lines and vice versa. In
the case $\Lambda _{z}=0$ that will be considered below, crossings are
defined by
\begin{eqnarray}
&&W_{m_{I}^{(+)},m_{I}^{(-)}}^{(\mathrm{cross})}=\left(
m_{I}^{(-)}-m_{I}^{(+)}\right) F\equiv kF,\qquad  \notag \\
&&k=m_{I}^{(-)}-m_{I}^{(+)}=-2I,-2I+1,\ldots ,2I.  \label{WCrossmmpr}
\end{eqnarray}
There are total $4I+1$ crossings. For a given $k$, the values of $%
m_{I}^{(-)} $ and $m_{I}^{(+)}$ satisfy
\begin{equation}
m_{I}^{(-)}-m_{I}^{(+)}=k,\qquad -I\leq m_{I}^{(-)},m_{I}^{(+)}\leq I.
\end{equation}
For the leftmost crossing with $k=-2I$ there is only one solution, $\left(
m_{I}^{(-)}=-I,m_{I}^{(+)}=I\right) ,$ i.e., only one pair of lines is
crossing here. Similarly, for the rightmost crossing with $k=2I$ there is
only one pair of crossing lines with $\left(
m_{I}^{(-)}=I,m_{I}^{(+)}=-I\right) .$ For the crossing with $k=-2I+1,$
there are two pairs of crossing lines with $\left(
m_{I}^{(-)}=-I+1,m_{I}^{(+)}=I\right) $ and $\left(
m_{I}^{(-)}=-I,m_{I}^{(+)}=I-1\right) .$ For $k=0$ there are $I$ crossing
lines with all possible values of $m_{I}$ and the corresponding $%
m_{I}^{(+)}=m_{I}.$ In general, the allowed values of $m_{I}^{(-)}$ and $%
m_{I}^{(+)}$ are in the intervals
\begin{equation}
\max \left( -I-k,-I\right) \leq m_{I}^{(+)}\leq \min (I-k,I).
\label{mplusInterval}
\end{equation}
and
\begin{equation}
\max \left( -I+k,-I\right) \leq m_{I}^{(-)}\leq \min (I+k,I)
\label{mminusInterval}
\end{equation}

In the case $\Delta \ll A,\Lambda _{x}$ crossings are well separated from
each other. In this case the process consists of tunneling transitions at
the crossings and free evolution with phase accumulation in the ranges
between them. Transition at each crossing is described by the LZ scattering
matrix
\begin{equation}
M(\Delta )=\left(
\begin{array}{cc}
\sqrt{P} & \mathrm{sign}(\Delta )\sqrt{1-P}e^{-i\phi } \\
-\mathrm{sign}(\Delta )\sqrt{1-P}e^{i\phi } & \sqrt{P}
\end{array}
\right) ,  \label{MDef}
\end{equation}
where $P$ given by Eq.\ (\ref{PLZ}) is the Landau-Zener staying probability
and
\begin{equation}
\phi =\pi /4+\mathrm{Arg}\Gamma \left( 1-i\delta \right) +\delta \left( \ln
\delta -1\right)  \label{MphiDef}
\end{equation}
with $\delta \equiv \varepsilon /(2\pi )$ is the scattering phase and $%
\Gamma $ is Gamma function.

Evolution of the wave function between level crossings reduces to the
accumulation of the phase factors $\exp \left[ i\Phi _{\xi ,m_{I}}(t)\right]
,$ where the phases are given by
\begin{equation}
\Phi _{\xi ,m_{I}}(t)=-\frac{1}{\hbar }\int^{t}dt^{\prime }E_{\xi
,m_{I}}(t^{\prime }).  \label{Phase}
\end{equation}
The change of the phase between the $k$th and $\left( k+1\right) $th
crossings is given by
\begin{eqnarray}
\Phi _{\xi ,m_{I}}^{(k+1/2)} &=&\frac{1}{\hbar }\int_{t_{k}}^{t_{k+1}}dt^{%
\prime }\left[ \frac{\xi }{2}vt+Fm_{I}\right]  \notag \\
&=&\frac{1}{\hbar }\frac{F}{v}\left[ \frac{\xi }{4}F\left( 2k+1\right)
+Fm_{I}\right]
\end{eqnarray}
or, finally,
\begin{equation}
\Phi _{\xi ,m_{I}}^{(k+1/2)}=\frac{\varepsilon F^{2}}{\pi \Delta ^{2}}\left[
\xi \left( k+\frac{1}{2}\right) +2m_{I}\right] .  \label{Phikkplusone}
\end{equation}

Evolution of the state on the interval between the $k$th and $\left(
k+1\right) $th crossings is given by
\begin{eqnarray}
c_{\xi ,m_{I}}^{\mathrm{out}} &=&T_{\xi ,m_{I};\xi ^{\prime },m_{I}^{\prime
}}^{(k+1/2)}c_{\xi ^{\prime },m_{I}^{\prime }}^{\mathrm{in}}  \notag \\
&=&\exp \left[ i\Phi _{\xi ,m_{I}}^{(k+1/2)}\right] \delta _{\xi ^{\prime
}\xi }\delta _{m_{I}^{\prime }m_{I}}c_{\xi ^{\prime },m_{I}^{\prime }}^{%
\mathrm{in}}
\end{eqnarray}
or in the vector-matrix form
\begin{equation}
\mathbf{c}^{\mathrm{out}}=\mathbb{T}^{(k+1/2)}\mathbf{c}^{\mathrm{in}}.
\end{equation}

The change of the state across the $k$th resonance is described by
\begin{equation}
c_{\xi ,m_{I}}^{\mathrm{out}}=A_{\xi ,m_{I};\xi ^{\prime },m_{I}^{\prime
}}^{(k)}c_{\xi ^{\prime },m_{I}^{\prime }}^{\mathrm{in}},  \label{cTransAk}
\end{equation}
or
\begin{equation}
\mathbf{c}^{\mathrm{out}}=\mathbb{A}^{(k)}\mathbf{c}^{\mathrm{in}},
\end{equation}
where for $\xi ,m_{I}$ and $\xi ^{\prime },m_{I}^{\prime }$ describing a
pair of crossing levels $A_{\xi ,m_{I};\xi ^{\prime },m_{I}^{\prime }}^{(k)}$
is given by the LZ scattering matrix $M$, otherwise, if there is no
crossing, it is $A_{\xi ,m_{I};\xi ^{\prime },m_{I}^{\prime }}^{(k)}=\delta
_{\xi ,\xi ^{\prime }}\delta _{m_{I}^{\prime }m_{I}}.$ To formulate the
condition of crossing, one has to consider the cases $\xi ^{\prime }=\pm 1$
separately. For $\xi ^{\prime }=+1$ (descending line) \ the nuclear quantum
number $m_{I}^{\prime }=m_{I}^{(+)}$ should be in the interval given by Eq.\
(\ref{mplusInterval}), for a crossing to be realized, then in the scattered
state one has $m_{I}=m_{I}^{(-)}=m_{I}^{\prime }+k.$ For $\xi ^{\prime }=-1$
(ascending line) \ the nuclear quantum number $m_{I}^{\prime }=m_{I}^{(-)}$
should be in the interval given by Eq.\ (\ref{mminusInterval}), for a
crossing to be realized, then in the scattered state one has $%
m_{I}=m_{I}^{(+)}=m_{I}^{\prime }-k.$ Thus in the general case one obtains
\begin{eqnarray}
A_{\xi ,m_{I};\xi ^{\prime },m_{I}^{\prime }}^{(k)} &=&M_{\xi ,\xi ^{\prime
}}(\Delta _{\xi ,m_{I},m_{I}-k})  \notag \\
&&\times \left( \delta _{\xi \xi ^{\prime }}\delta _{m_{I}m_{I}^{\prime
}}+\delta _{\xi ,-\xi ^{\prime }}\delta _{m_{I},m_{I}^{\prime }+k}\right)
\end{eqnarray}
for $\xi ^{\prime }=+1,\,\max \left( -I-k,-I\right) \leq m_{I}^{\prime }\leq
\min (I-k,I),$%
\begin{eqnarray}
A_{\xi ,m_{I};\xi ^{\prime },m_{I}^{\prime }}^{(k)} &=&M_{\xi ,\xi ^{\prime
}}(\Delta _{\xi ,m_{I},m_{I}+k})  \notag \\
&&\times \left( \delta _{\xi \xi ^{\prime }}\delta _{m_{I}m_{I}^{\prime
}}+\delta _{\xi ,-\xi ^{\prime }}\delta _{m_{I},m_{I}^{\prime }-k}\right)
\end{eqnarray}
for $\xi ^{\prime }=-1,\,\max \left( -I+k,-I\right) \leq m_{I}^{\prime }\leq
\min (I+k,I),$ and $A_{\xi ,m_{I};\xi ^{\prime },m_{I}^{\prime
}}^{(k)}=\delta _{\xi ,\xi ^{\prime }}\delta _{m_{I}^{\prime }m_{I}}$ in all
other cases.

The evolution of the state across the whole grid is given by
\begin{eqnarray}
\mathbf{c}^{\mathrm{out}} &=&\mathbb{A}^{(2I)}\mathbb{T}^{(2I-1/2)}\mathbb{A}%
^{(2I-1)}\cdots  \notag \\
&&\times \mathbb{A}^{(-2I+1)}\mathbb{T}^{(-2I+1/2)}\mathbb{A}^{(-2I)}\mathbf{%
c}^{\mathrm{in}},  \label{Fullcout}
\end{eqnarray}
where $\mathbf{c}^{\mathrm{in}}$ is the state of the system before the
leftmost crossing. The calculation can be made recurrent introducing the
scattering matrix after the $k$th crossing:
\begin{equation}
\mathbf{c}_{k}^{\mathrm{out}}=\mathbb{L}^{(k)}\mathbf{c}^{\mathrm{in}},
\end{equation}
where $\mathbf{c}^{\mathrm{in}}$ is the initial state, same as in Eq.\ (\ref
{Fullcout}). One has $\mathbb{L}^{(-2I)}=\mathbb{A}^{(-2I)}$
\begin{equation}
\mathbb{L}^{(k)}=\mathbb{A}^{(k)}\mathbb{T}^{(k-1/2)}\mathbb{L}^{(k-1)},
\end{equation}
$k=-2I+1,-2I+1,\ldots ,2I.$ In the final state it is
\begin{equation}
\mathbf{c}^{\mathrm{out}}=\mathbf{c}_{2I}^{\mathrm{out}}=\mathbb{L}^{(2I)}%
\mathbf{c}^{\mathrm{in}},
\end{equation}
$\mathbb{L}^{(2I)}$ being the full scattering matrix of the grid.

If in the initial state the nuclear spin is in its ground state, application
of Eq.\ (\ref{Fullcout}) and calculation of the staying probability yields
the result of Eq.\ (\ref{PLZ}), no effect of nuclear spins. For $I=1/2$ the
problem can be solved analytically. If the nuclear spin is in its excited
state, then the result is
\begin{eqnarray}
P &=&e^{-\varepsilon }+\exp \left( -\frac{\varepsilon A^{2}}{F^{2}}\right) %
\left[ 1-\exp \left( -\frac{\varepsilon A^{2}}{F^{2}}\right) \right]  \notag
\\
&&\times \left[ 1-\exp \left( -\frac{\varepsilon \Lambda _{x}^{2}}{F^{2}}%
\right) \right] 4\sin ^{2}\frac{\varepsilon F^{2}}{\pi \Delta ^{2}}.
\label{PinfIonehalf}
\end{eqnarray}
In the fast-sweep limit $\varepsilon \ll 1$ this formula simplifies to
\begin{equation}
P\cong e^{-\varepsilon }+\varepsilon ^{2}\left( \frac{A\Lambda _{x}}{F^{2}}%
\right) ^{2}4\sin ^{2}\frac{\varepsilon F^{2}}{\pi \Delta ^{2}},
\label{PinfIonehalfsmalleps}
\end{equation}
i.e., at the linear order in $\varepsilon $ Eq.\ (\ref{PLZ}) is robust. For $%
A\ll \Lambda _{x}$ the envelope of Eq. (\ref{PinfIonehalf}) has a local
maximum at $\varepsilon \sim \Lambda _{x}^{2}/A^{2}$ with $P_{\max }\sim 1.$
The asymptotic $\varepsilon \gg 1$ behavior Eq.\ (\ref{PinfIonehalf}) is
determined by the factor $\exp \left( -\varepsilon A^{2}/F^{2}\right) $ that
is slowly decaying for $A\ll \Lambda _{x}.$ The factor $A^{2}/F^{2}$ in the
exponential is the square of the antidiagonal elements of \ the matrix in
Eq.\ (\ref{VMatrOneHalf}) and it corresponds to the splitting that is much
smaller than $\Delta .$ To the contrary, for $A\gg \Lambda _{x}$ the result
is close to $P=e^{-\varepsilon }.$ This means that the influence of nuclear
spins strongly coupled to the electronic spin is small.

In fact, analytical expressions, although too cumbersome, can be obtained
for any $I.$ The common feature of the solutions for all $I$ is a standard
decay of $P(\varepsilon )$ for $\varepsilon \gg 1$ in the case $A\gg \Lambda
_{x}$ and an extremely slow decay in the case $A\ll \Lambda _{x}.$ This
becomes clear from the analysis of the LZ grids in both limiting cases,
shown in Fig.\ \ref{Fig-LZ-grid}. For $A\gg \Lambda _{x}$ the large
splittings in Fig.\ \ref{Fig-LZ-grid} (a) line up horizontally, and in the
slow-sweep limit the system cannot cross the dotted line moving along the
ascending levels from left to right. It follows the exact levels and
adiabatically turns down to the descending levels that leads to a full LZ
transition, $P\approx 0.$ On the contrary, for $A\ll \Lambda _{x}$ the large
splittings in Fig.\ \ref{Fig-LZ-grid} (b) line up vertically. Although at
slow sweep transitions at these crossings are adiabatic, there are smaller
crossings at $W<0$ and $W>0$ where transitions are non-adiabatic, so that
the electronic spin can end up in its both states. As a result, $%
P(\varepsilon )$ vanishes only at extremely slow sweep, $\varepsilon \ggg 1$%
, at which the smallest crossings become adiabatic.

\begin{figure}[t]
\unitlength1cm
\begin{picture}(11,5.5)
\centerline{\psfig{file=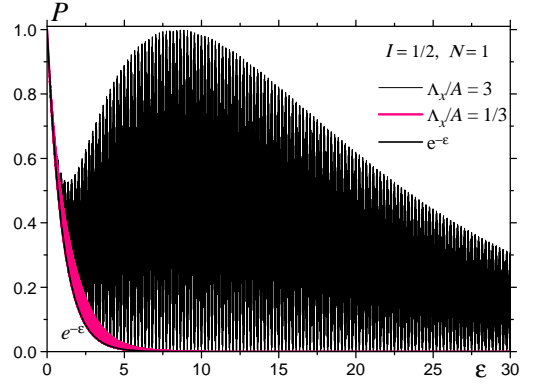,angle=-90,width=8cm}}
\end{picture}
\caption{Asymptotic staying probability $P$ for an electronic spin coupled
to one nuclear spin $I=1/2,$ initially in its excited state. The influence
of strongly coupled nuclear spins becomes small, the curve $\Lambda
_{x}/A=1/3.$}
\label{Fig-Ionehalf-N1}
\end{figure}
\begin{figure}[t]
\unitlength1cm
\begin{picture}(11,5.5)
\centerline{\psfig{file=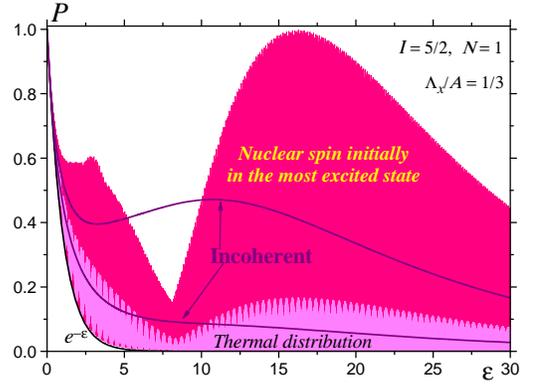,angle=-90,width=8cm}}
\end{picture}
\caption{$P$ for one nuclear spin $I=5/2,$ with $\Lambda _{x}/A=1/3.$ Effect
of the nuclear spin is large even for a small ratio $\Lambda _{x}/A$ due to
large $I.$ Averaging over the initial thermal distribution of nuclear spins (%
$T=\infty )$ reduces the effect. The incoherent approximation is defined in
the next section.}
\label{Fig-Ifivehalf-N1}
\end{figure}

Next we present figures illustrating the dependence of the asymptotic
staying probability $P$ on the sweep-rate parameter $\varepsilon $ for the
model with one nuclear spin. Fig.\ \ref{Fig-Ionehalf-N1} shows the results
for $I=1/2$ and different ratios between $\Lambda _{x}$ and the
electron-nuclear coupling $A.$ The effect of nuclear spins is large for $%
\Lambda _{x}\gtrsim A,$ in accordance with Eq.\ (\ref{PinfIonehalf}).

With increasing $I$ the number of crossings in the LZ grid and thus the
effect of the coupling to the nuclear spin on $P$ increases, as one can see
in Fig.\ \ref{Fig-Ifivehalf-N1}. On the other hand, the effect is maximal
for the nuclear spin initially in the mostly excited state and decreases in
the case of the thermal distribution ($T=\infty )$ in the initial state. For
$I=5/2$ the effect is large even for $\Lambda _{x}/A=1/3$. This suggests
that $\Lambda _{x}/A$ is not a proper parameter to describe relative
strength of different terms in the Hamiltonian. As commented upon below Eq.\
(\ref{DeltammLamSmall}), the ratio between the sub-anti-diagonal to the
dominant anti-diagonal terms is
\begin{equation}
l_{m,m+1}\Lambda _{x}/A\sim I\Lambda _{x}/A  \label{LamxRatio}
\end{equation}
that is close to 1 in Fig.\ \ref{Fig-Ifivehalf-N1}. Still in Mn$_{12}$ the
hyperfine coupling $A$ is very strong, see comment below Eq.\ (\ref{HamSI}).
External transverse field of 2 T creates $\Lambda _{x}/A\simeq 2/40=0.05,$
so that with $I=5/2$ one has $I\Lambda _{x}/A\simeq 0.12.$ Since this
parameter enters squared the effective sweep parameter $\varepsilon
_{mm^{\prime }}$ at sub-primary crossings, the direct effect of Mn$^{55}$
spins on the LZ transitions in Mn$_{12}$ should be small for not too strong
magnetic fields. The physical reason for this is that nuclear spins do not
have a sufficient dynamics to undergo transitions together with the
electronic spin. The only thing that they can do is to create a static bias
on the electronic spin, random because of the thermal distribution of
nuclear spins. The latter becomes very important in combination with the
dipole-dipole interaction between the electronic spins.\cite{garsch05prb}

\section{Many nuclear spins and incoherent approximation}

\label{Sec-Many-incoh}

\begin{figure}[t]
\unitlength1cm
\begin{picture}(11,5.5)
\centerline{\psfig{file=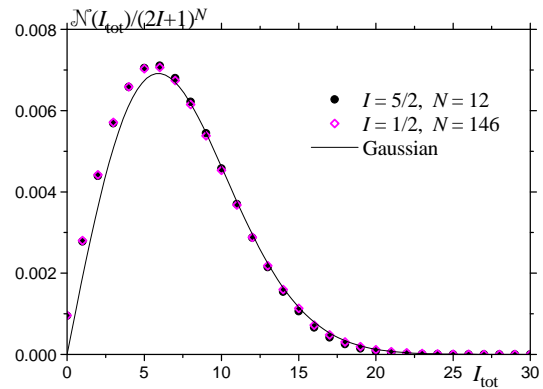,angle=-90,width=8cm}}
\end{picture}
\caption{Normalized distribution function for the total nuclear spin $I_{%
\mathrm{tot}}.$ }
\label{Fig-NItot}
\end{figure}

Let us now consider the model of many nuclear spins coupled to the
electronic spin. The simplest way to generalize the results of the preceding
section is to assume the equivalence of nuclear spins and the same coupling
to the central electronic spin. In this case Eq.\ (\ref{HamSI}) is replaced
by
\begin{equation}
\hat{H}=-\frac{1}{2}W(t)\sigma _{z}-\frac{1}{2}\Delta \sigma _{x}-\left(
A\sigma _{z}+\Lambda _{z}\right) I_{\mathrm{tot,}z}-\Lambda _{x}I_{\mathrm{%
tot,}x},  \label{HamSIN1}
\end{equation}
where $\mathbf{I}_{\mathrm{tot}}=\sum_{i=1}^{N}\mathbf{I}_{i}$ is the total
nuclear spin. In fact, 12 nuclear spins in Mn$_{12}$ couple to the atomic
spins with somewhat different coupling constants\cite{harpolvil96} but this
will be ignored for a\ moment. The length of the total nuclear spin is
dynamically conserved, $\left[ \hat{H},\left( \mathbf{I}_{\mathrm{tot}%
}\right) ^{2}\right] =0.$ Thus the states of the nuclear subsystem can be
classified by the value of the total spin $I_{\mathrm{tot}}\leq NI$ and its
projection $m_{I}$ on some axis. This reduces the problem to the one studied
in the preceding section. In particular, for all nuclear spins in the ground
state the nuclear subsystem will evolve as a single spin $I_{\mathrm{tot}%
}=NI.$ In the realistic case of thermal distribution of nuclear spins one
has to average the solutions for particular values of $I_{\mathrm{tot}}$ and
$m_{I}$ in the initial state over the distrubution of $I_{\mathrm{tot}}$ and
$m_{I}.$

The number of realizations $\mathcal{N}(I_{\mathrm{tot}})$ of $I_{\mathrm{tot%
}}$ can be computed recurrently.\cite{garchusch00prb} If the total spin of a
system of $N$ nuclei is $I_{\mathrm{tot}}$, the total spin of its subsystem
of $N-1$ nuclei $I_{\mathrm{tot}}^{\prime }$ assumes the values $|I_{\mathrm{%
tot}}-I|\leq I_{\mathrm{tot}}^{\prime }\leq \min \{I_{\mathrm{tot}%
}+I,(N-1)I\}$. Thus for the number of realizations $\mathcal{N}(I_{\mathrm{%
tot}},N)$ one can write
\begin{equation}
\mathcal{N}(I_{\mathrm{tot}},N)=\sum_{I_{\mathrm{tot}}^{\prime }=|I_{\mathrm{%
tot}}-I|}^{\min \{I_{\mathrm{tot}}+I,(N-1)I\}}\mathcal{N}(I_{\mathrm{tot}%
}^{\prime },N-1).  \label{Recurr}
\end{equation}
The initial condition for this recurrence relation is $\mathcal{N}(I_{%
\mathrm{tot}}^{\prime },2)=1$ for $0\leq I_{\mathrm{tot}}^{\prime }\leq 2I$.
The quantity $\mathcal{N}(I_{\mathrm{tot}})$ obeys the normalization
condition
\begin{equation}
\sum_{I_{\mathrm{tot}}=\mathrm{frac}(NI)}^{NI}(2I_{\mathrm{tot}}+1)\mathcal{N%
}(I_{\mathrm{tot}})=(2I+1)^{N}.  \label{NNorm}
\end{equation}
For $NI\gg 1$, the quantity $(2I_{\mathrm{tot}}+1)\mathcal{N}(I_{\mathrm{tot}%
})/(2I+1)^{N}$ is the high-temperature distribution function of the
magnitude of $I_{\mathrm{tot}}$ and it is well approximated by $4\pi I_{%
\mathrm{tot}}^{2}F(\mathbf{I}_{\mathrm{tot}})$, where $F(\mathbf{I}_{\mathrm{%
tot}})$ is a normalized Gaussian function with respect to the three
components of $\mathbf{I}_{\mathrm{tot}}$. \cite{garchu97prb} Thus for the
asymptotic $NI\gg 1$ form of the distribution function of $I_{\mathrm{tot}}$
normalized by 1 one has
\begin{equation}
\frac{\mathcal{N}(I_{\mathrm{tot}})}{(2I+1)^{N}}\cong \frac{2\pi I_{\mathrm{%
tot}}}{(2\pi \sigma _{I})^{3/2}}\exp \left( -\frac{I_{\mathrm{tot}}^{2}}{%
2\sigma _{I}}\right) .  \label{NAsymp}
\end{equation}
where $\sigma _{I}=(N/3)I(I+1)$. It has a maximum at $I_{\mathrm{tot}}=\sqrt{%
\sigma _{I}}$ which is about 6 for Mn$_{12}$ ($I=5/2$, $N=12$). Fig. \ref
{Fig-NItot} shows an agreement between the exactly computed $\mathcal{N}(I_{%
\mathrm{tot}})/(2I+1)^{N}$ and its Gaussian approximation for $I=5/2$ and $%
N=12$. This agreement improves for higher values of $NI$. One can see that
practically the same result for $\mathcal{N}(I_{\mathrm{tot}})/(2I+1)^{N}$
can be achieved with $I=1/2$ and $N=146$. Thus in the model described by
Eq.\ (\ref{HamSIN1}), the effect of 12 Mn$^{55}$ spins is the same as that
of $\ $about $150$ protons.

\begin{figure}[t]
\unitlength1cm
\begin{picture}(11,5.5)
\centerline{\psfig{file=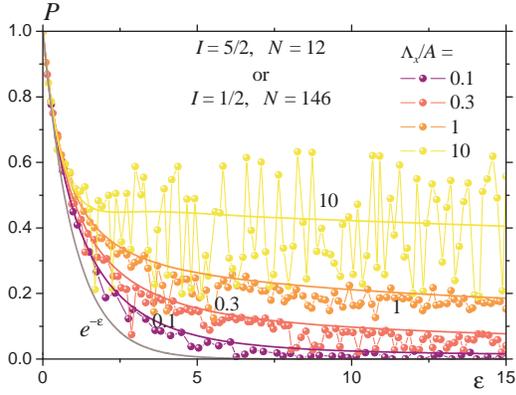,angle=-90,width=8cm}}
\end{picture}
\caption{$P(\protect\varepsilon )$ for the electronic spin coupled to $N=12$
nuclear spins $I=5/2$ (initially in a thermal state with $T=\infty $) for
different ratios $\Lambda _{x}/A$ and $\Lambda _{z}=0.$ Results of the
incoherent approximation are shown by solid lines of same color.}
\label{Fig-Coh_vs_incoh}
\end{figure}

Summation over different values of $I_{\mathrm{tot}}$ makes calculations
lengthier. In addition, there are high values of $I_{\mathrm{tot}}$ with a
small statistical weight that consume a lot of time but do not change the
result. Thus one has to introduce a cut-off on $I_{\mathrm{tot}}.$ The
results for $I=5/2$ and $N=12$ (or for $I=1/2$ and $N=146$) and $\Lambda
_{z}=0$ are shown in Fig.\ \ref{Fig-Coh_vs_incoh}. Since the most probable
value of $I_{\mathrm{tot}}$ in this case is between 5 and 6, deviations from
the standard LZ effect are noticeable for $\Lambda _{x}/A$ down to 0.1 (see
discussion at the end of preceding section).

\begin{figure}[t]
\unitlength1cm
\begin{picture}(11,5.5)
\centerline{\psfig{file=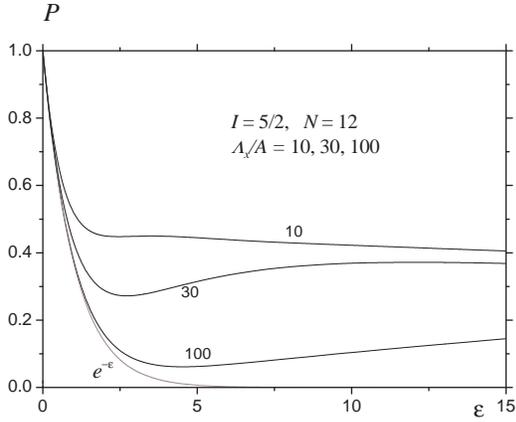,angle=-90,width=8cm}}
\end{picture}
\caption{$P(\protect\varepsilon )$ for the electronic spin coupled to $N=12$
nuclear spins $I=5/2$ for large ratios $\Lambda _{x}/A$ and $\Lambda _{z}=0$
within the incoherent approximation.}
\label{Fig-Peps-incoh_large-rat}
\end{figure}

One can see in Fig.\ \ref{Fig-Coh_vs_incoh} that summations over both $I_{%
\mathrm{tot}}$ and $m_{I}$ in the initial state lead to decreasing of
quantum-mechanical oscillations in $P(\varepsilon ).$ However, averaging out
the oscillations is incomplete because for all values of $I_{\mathrm{tot}}$
the $W$ intervals in the LZ grid are the same, although the number of
crossings change. In real situations the oscillations should be averaged
out, mainly because of different couplings $A$ to different nuclear spins
that increases the number of crossings from $\left( 2I_{\mathrm{tot}%
}+1\right) ^{2}$ to $(2I+1)^{2N}$ . To take the phase averaging into account
without increasing the size of the problem, one can use the incoherent
approximation. In the case of well-separated resonances one can consider the
occupation numbers of the states
\begin{equation}
p_{\xi ,m_{I}}=\left| c_{\xi ,m_{I}}\right| ^{2}.
\end{equation}
The change of the state at the $k$th resonance is described by Eq.\ (\ref
{cTransAk}). Then for the probabilities one has
\begin{equation}
p_{\xi ,m_{I}}^{\mathrm{out}}=A_{\xi ,m_{I};\xi ^{\prime },m_{I}^{\prime
}}^{(k)}A_{\xi ,m_{I};\xi ^{\prime \prime },m_{I}^{\prime \prime }}^{(k)\ast
}c_{\xi ^{\prime },m_{I}^{\prime }}^{\mathrm{in}}c_{\xi ^{\prime \prime
},m_{I}^{\prime \prime }}^{\mathrm{in\ast }}.
\end{equation}
Neglecting interference effects or, in other words, averaging over phases
amounts to the aproximation
\begin{equation}
c_{\xi ^{\prime },m_{I}^{\prime }}^{\mathrm{in}}c_{\xi ^{\prime \prime
},m_{I}^{\prime \prime }}^{\mathrm{in\ast }}\Rightarrow p_{\xi ^{\prime
},m_{I}^{\prime }}\delta _{\xi ^{\prime }\xi ^{\prime \prime }}\delta
_{m_{I}^{\prime }m_{I}^{\prime \prime }}.
\end{equation}
Then the change of the occupation numbers across a resonance is described by
\begin{equation}
p_{\xi ,m_{I}}^{\mathrm{out}}=B_{\xi ,m_{I};\xi ^{\prime },m_{I}^{\prime
}}^{(k)}p_{\xi ^{\prime },m_{I}^{\prime }}^{\mathrm{in}},  \label{pTrans}
\end{equation}
where
\begin{equation}
B_{\xi ,m_{I};\xi ^{\prime },m_{I}^{\prime }}^{(k)}=\left| A_{\xi ,m_{I};\xi
^{\prime },m_{I}^{\prime }}^{(k)}\right| ^{2}.  \label{BDef}
\end{equation}
Incoherent approximation requires only a slight change of the computational
method. Instead of Eq.\ (\ref{Fullcout}) one has
\begin{equation}
\mathbf{p}^{\mathrm{out}}=\mathbb{B}^{(2I)}\mathbb{B}^{(2I-1)}\cdots \mathbb{%
B}^{(-2I+1)}\mathbb{B}^{(-2I)}\mathbf{p}^{\mathrm{in}}.
\end{equation}
In particular, for one nuclear spin $I=1/2$ one obtains Eq.\ (\ref
{PinfIonehalf}) with $\sin ^{2}(\ldots )\Rightarrow 1/2.$ \

\begin{figure}[t]
\unitlength1cm
\begin{picture}(11,5.5)
\centerline{\psfig{file=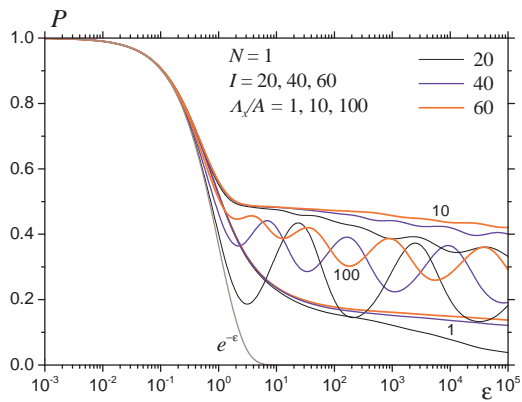,angle=-90,width=8cm}}
\end{picture}
\caption{$P(\protect\varepsilon )$ for the electronic spin coupled to one
large ``nuclear spin'' for different ratios $\Lambda _{x}/A$ and $\Lambda
_{z}=0$ within the incoherent approximation. }
\label{Fig-Peps-incoh_log-eps}
\end{figure}

Results of the incoherent approximation are shown in Figs.\ \ref
{Fig-Ifivehalf-N1} and \ref{Fig-Coh_vs_incoh} by smooth curves. The results
similar to those in Fig.\ \ref{Fig-Coh_vs_incoh} but for large ratios $%
\Lambda _{x}/A$ are shown in Fig.\ \ref{Fig-Peps-incoh_large-rat} within the
incoherent approximation. For very large $\Lambda _{x}/A,$ the sub-primary
splittings [left and right from the primary splittings at $W=0$ in Fig.\ \ref
{Fig-LZ-grid} (b)] are very small, so that for not too slow sweep
transitions occur at $W=0$ and they become adiabatic with $P\approx 0$ for $%
\varepsilon \gtrsim 1.$ However, with further increase of $\varepsilon $
transitions at crossings with sub-primary splittings begin, and the curve $%
P(\varepsilon )$ goes up again. Fig.\ \ref{Fig-Peps-incoh_log-eps} gives an
idea of the dependence $P(\varepsilon )$ for a large nuclear spin. The
apparent overall dependence is $P(\varepsilon )\simeq C_{1}-C_{2}\log
\varepsilon ,$ seen in Fig.\ \ref{Fig-Peps-incoh_log-eps} over several
decades in $\varepsilon .$ There are slow oscillations for $\Lambda
_{x}/A=100$ that tend to disappear for large $I.$

\section{Discussion}

The model of a two-level system coupled to one or many environmental spins
considered above is one of basic models illustrating many-body Landau-Zener
effect with many avoided level crossings forming a Landau-Zener grid. Exact
results of the multilevel LZ effect such as the no-go theorem and DOBE
formula apply to this model. In particular, if the environmental spins are
initially in their ground states, the asymptotic staying probability is
unchanged and given by Eq.\ (\ref{PLZ}). Accordingly no isotope effect
exists in this case. However, such an effect occurs when preparing nuclear
spins e.g., in a thermal ensemble.

The model under consideration is relevant for spin tunneling in molecular
magnets, with coupling to nuclear spins, although comparison with
experimental results requires taking into account the dipole-dipole
interaction between electronic spins at the same time. The latter is more
difficult and up to now could be done only in the fast-sweep limit $%
\varepsilon \ll 1.\cite{garsch05prb}$

Surprisingly, nuclear spins strongly coupled to the electronic spin, such as
12 Mn$^{55}$ nuclear spins $I=5/2$ in Mn$_{12},$ do not significantly change
the asymptotic survival probability $P(\varepsilon )$ for the electronic
spin. This follows from the analysis of the Landau-Zener grid, Fig.\ \ref
{Fig-LZ-grid} (a). In physical terms, nuclear spins in the case $\Lambda
_{x}\ll A$ do not have a sufficient dynamics to undergo transitions together
with the electronic spin. They rather act on the latter as static random
fields that have no effect by themselves but do significantly reduce the
influence of the DDI on the LZ effect.$\cite{garsch05prb}$

To the contrast, protons in molecular magnets can influence the LZ effect
directly. Although the nuclear magneton $\mu _{n}$ is small, the energy of
the dipole-dipole interaction with nuclear spins can exceed tunnel splitting
$\Delta .$ For instance, the ground-state splitting in Fe$_{\text{8}}$ in
zero field is only $\Delta /k_{\mathrm{B}}\simeq 10^{-7}$ K. The energy bias
on the electron spin from a proton at distance $r$ is given by $W\sim 2Sg\mu
_{\mathrm{B}}\mu _{p}/r^{3},$ where $g=2$ and $\mu _{p}=2.79\mu _{n}.$ One
has $\Delta \sim W$ at the distance $r_{\Delta }\simeq 72$ \AA . The volume
of a corresponding sphere comprises about 400 unit cells of a molecular
magnet. As each molecule, occupying a unit cell, comprises about 120
protons, \cite{weretal00prl} the number of protons creating bias $W\gtrsim
\Delta $ is about $N=50000$. These protons are typically weakly coupled to
the electronic spin $\left( A\ll \Lambda _{x}\right) $ since their Zeeman
interaction with the external field greatly exceeds their NDDI with
electronic spins. Applying an external magnetic field, one can drastically
increase $\Delta $ and thus reduce $N.$

Effect of protons in principle could be tackled with the method described in
Sec.\ \ref{Sec-General}. However, as each proton couples to the electronic
spin with its own coupling constant $A,$ the number of lines in the LZ grid,
$2^{N},$ is too large to implement a working computational algorithm. The
first expected effect of so many crossings is complete averaging out of
quantum-mechanical phase oscillations that are still seen in Fig.\ \ref
{Fig-Coh_vs_incoh}. This effect can easily be accounted for by the
incoherent approximation of Sec.\ \ref{Sec-Many-incoh}. After that different
couplings $A$ become less important, and one can get a qualitative idea of
the effect of protons from the analysis in the preceding section. The most
probable combined nuclear spin $I_{\mathrm{tot}}$ in the distribution of
Eq.\ (\ref{NAsymp}) is $\sqrt{\sigma _{I}}\approx 110$ that is of the same
order as the large ``nuclear spins'' simulated in Fig.\ \ref
{Fig-Peps-incoh_log-eps}. The resulting $P(\varepsilon )$ is decaying
extremely slowly because of a strong non-adiabaticity at large $\varepsilon $
induced by weak avoided crossings outside the ellipse in Fig.\ \ref
{Fig-Levels-Large_spin}. As the width of the distribution of $I_{\mathrm{tot}%
}$ is of the same order as the most probable value of $I_{\mathrm{tot}},$
slow oscillations in Fig.\ \ref{Fig-Levels-Large_spin} should be averaged
out after summation over all $I_{\mathrm{tot}}$. One can see that the
results for $P(\varepsilon )$ in the case $A\ll \Lambda _{x}$ slowly
approach 1/2. This is in accord with the results of Ref.\ %
\onlinecite{kay93prb} where the influence of environment was modeled by a
density-matrix equation with dephasing.

It does not make sense, however, to further elaborate on the effect of
protons in this paper because in the practical case $A\ll \Lambda _{x}$
coupling to electronic spins only slightly perturbs protons. This might be
an indication of a possibility to solve the problem by another and more
efficient method.

The work by D. A. G. was supported by the NSF grant No. DMR-0703639.

\bibliographystyle{prsty}
\bibliography{gar-oldworks,gar-books,gar-own,gar-nuc-spin,gar-tunneling,gar-lz,gar-relaxation,chu-own}

\end{document}